# Investigation of high frequency transducers and coded signals suitable for cartilage volume imaging

Abhishek Ranjan[1], Chengxiang peng[1], Anowarul Habib[1], Sanat Wagle[2], Frank Melandsø[1]
[1]Department of Physics and Technology, UiT The Arctic University of Norway, Tromsø, Norway
[2]Elop AS, Nordvikvegen 50, 2316 Hamar, Norway
Email- frank.melandso@uit.no

*Abstract:* Cartilage degeneration in joints causing pain and various types of knee problems is a serious problem-affecting people in all ages. Degenerated articular cartilage is also known as a central hallmark of osteoarthritis, which is a complex musculoskeletal disorder involving numerous contributory genetic, constitutional and biomechanical factors. As a part of the cartilage degeneration, the volume occupied by the collagen fibers becomes reduced and the cell (chondrocyte) volume increased. Since high frequency ultrasound has the capability of resolving individual chondrocytes, ultrasound has been suggested as a promising method for determining the cartilage status. In the current work, the main objective has been to compare images taken in vitro with different transducer types, in order to determine their suitability for cartilage imaging.

*Keywords:* *cartilage, acoustic microscopy, PVDF, transducers, LabVIEW*.

## I. INTRODUCTION

Scanning acoustic microscopy (SAM) is a wide field non-destructive and non-invasive technique that has been widely used over several decades for surface and subsurface microscopic imaging for industrial and biological specimens [1, 2]. SAM employs an acoustic energy which is transmitted via a coupling medium generally lying between 1 and 1000 MHz [3] The capabilities of SAM also includes the observation of internal structures, subsurface features, and structural characterization of materials and detecting changes in the elastic properties of solids [2, 4, 5].

The commercial transducers used in SAM are made up of ceramic, single crystals or thin films of piezoelectric materials. In the field of SAM, acoustic transducers having broad bandwidth, high sensitivity and mechanical flexibility has become crucial. Polyvinylidene difluoride (PVDF) and its copolymer is a piezoelectric material and inherently possess a number of benefits for acoustic transducer.
PVDF is a flexible material, which allows high degree of physical focusing without lenses [6]. PVDF transducers are also of significance because of their wide bandwidth, short impulse response [7]. PVDF and it copolymer films such as PVDF trifluoroethylene P(VDF-TrFE) offer an edge over conventional piezoelectric films, in terms of dielectric permittivity, higher phase velocity and higher electromechanical coupling. The copolymer P(VDF-TrFE) also has one of the highest piezoelectric activity (ratio 70:30) among known piezoelectric polymer materials [8].

The articular cartilage is a bonding tissue that acts as a cushion between joints. It is a hyaline, helping bones to glide over each other with low friction. It also lacks of blood vessels, lymphatics, and nerves [9-11]. The indication of degeneration of cartilage can be seen evidently from the changes in biomechanical properties. e.g. in rheumatoid arthritis or osteoarthritis (OA). OA is a progressive joint disease, involving gradual degradation of articular cartilage. The prevalence of OA can be viewed as an age-related phenomenon, as changes in the way chondrocytes function are often observed with growing age. The symptoms include pain, stiffness of the joint and weakness in muscles. Several morphological and structural modifications occur during the development of OA [12-15].

The clinical non-invasive imaging modalities such as radiography, magnetic resonance imaging, computed tomography, and conventional echography enable the detection of only severe cartilage lesions which is a characteristic of late stages of OA [16-18]. Since superficial cartilage degradation is the first sign of OA, its early detection would be of diagnostic importance. Acoustic microscopy, compared to other imaging modalities has the potential, to probe the surface or subsurface bioelastic properties of biological samples, thin films, circuits, piezoelectric materials etc. [19-22].We have fabricated a transducer made up from P (VDF-TrFe) material, which is adhesive free, having lower aperture diameter, and a lower f-number. The main purpose of the current work has therefore been to compare the performance of commercial and in-house transducers by high-frequency ultrasonic imaging of much thicker (mm-thick) cartilage samples.

## II. SAMPLE PREPARATION

Non-degenerated cartilage sample was obtained from the University Hospital of Northern Norway. The sample was from the patients (age between 50 to 60) who underwent a complete knee replacement. The patients having inflammatory joint diseases and advanced osteoarthritis were not included from the study. The study was approved by Regional Ethical Committee of Northern Norway (REK Nord 2014/920). A written consent was provided to all patients for the use of their cartilage for further study. We prepared the sample by slicing them carefully that resulted in a sample of around 1.25 mm thickness and then stored them in the PBS (Phosphate Buffered Saline) solution. The cartilage sample was placed inside a petri-dish (ibidis-50 mm) made up of polymer. The sample was fixed on the surface of the petri-dish using sticky tape. The petri-dish with cartilage was filled with water in order to facilitate a better acoustic coupling since air is a very poor transmitter of acoustic energy coming from transducers. Fig. 1 represents the optical image of a cartilage sample.

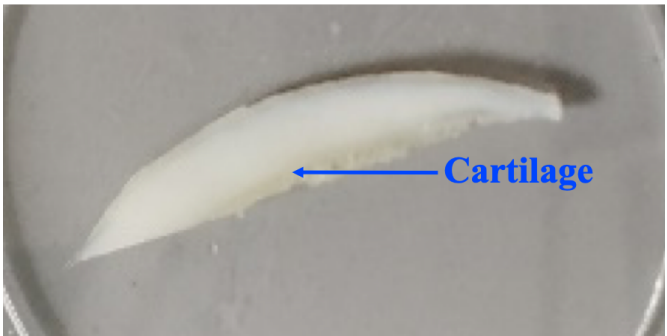

Fig. 1. Cartilage sample used in the experiments for acoustic imaging for both the in house fabricated and the commercial PVDF transducer.

### III. TRANSDUCER FABRICATION

The fabrication process of the P(VDF-TrFE) transducer begins with engraving of polymer polyethyleneimines (PEI) substrate having dimensions of 30×30 mm$^2$ and 0.85mm thickness with four spherical cavities of 2 mm diameter. After the milling process, the spherical cavities were cleaned with ethanol to remove unwanted material and grease. In order to increase the wettability of the substrate, plasma cleaner was employed. Sputtering was done through a high-resolution metal mask to get the first electrode (lower) layer. Subsequently, spin-coating was done on top of the patterned electrode fluid phase P(VDF-TrFE) (77:23, molar ratio) dissolved in appropriate amount of solvent. After that, the spin-coated substrate was degassed at 1mbar atm to vaporize the solvent and thereafter annealing was done at a temperature of 130° C for 8 hours to increase the crystallinity. Later on, the upper electrode was deposited silver with a thickness of ~80 nm on the top of P(VDF-TrFE) with the patterned metal mask. Each of the transducers in the substrate were connected directly onto a PCB using small spring contacts for poling and characterize the acoustic response. Fig. 2(a) represents the schematic diagram of the in-house P(VDF-TrFE) copolymer transducer, figure (b) shows the commercial transducer.

### IV. EXPERIMENTAL SETUP

The experiments were performed on an inverted microscope (Leica Dmi8) assembled with a custom-designed high precision scanning platform (ASI MS-2000) and other components. A detailed overview of the experimental setup can be found [11]. The scanning acoustic microscope was given ultrasonic functionality using FlexRIO modules and Programmable Gate Arrays (FPGA) hardware from National Instruments. This hardware constitutes of an arbitrary waveform generator (AT-1212), a 3W RF-amplifier (E&I 403LA) for pulse excitation, and a high-speed (1.6GS/s) 12-bit digitizer for pulse recording (NI-5772). All of the scanning operations including transducer movements and stage translations were controlled using LabVIEW. An ultrasonic transducer fixed above the stage is focussed onto the sample and the stage is scanned in raster mode. We used commercial and in-house fabricated transducers and the comparison based on different parameters are shown in table 1 below.

TABLE 1: In house fabricated PVDF and commercial PVDF transducers specification

| Transducer | Commercial PVDF | Fabricated P(VDF-TrFE) |
|---|---|---|
| Focal distance | 12.7mm | 2.6mm |
| Aperture Diameter | 6.35mm | 1.6mm |
| F-number | 2 | 1.625 |
| Central Frequency | 40 MHz | 40 MHz |

In order to explore the advantages of using coded waveforms for biological sample, both a Ricker wavelet and a long chirp-coded waveform were implemented on the imaging system. These waveforms were also investigated using both an in-house fabricated P(VDF-TrFE) transducer and a commercial PVDF transducer, both having center frequencies around 40 MHz

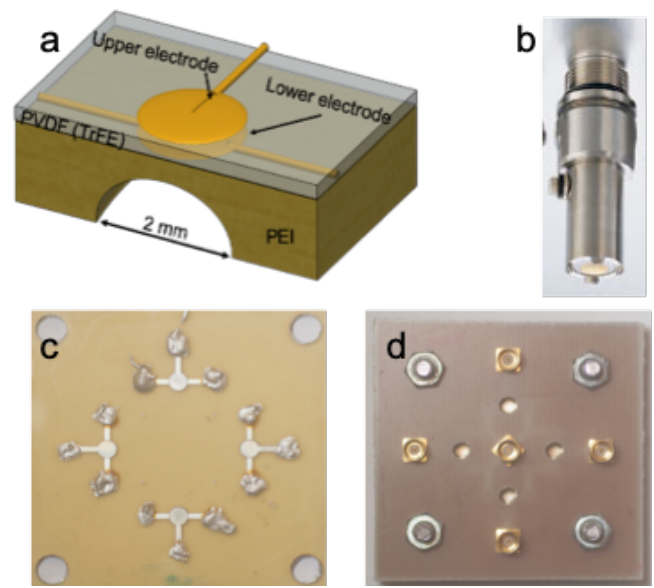

Fig. 2. (a) Drawing of the fabricated in-house P (VDF-TrFE) copolymer transducer, and (b) image of the commercial Olympus PVDF transducer, both with center frequencies around 40 MHz. (c) Optical image of a transducer panel containing 4 transducers, and (d) connection RF plugs for the transducers mounted on a printed circuit board.

Figs. 3 (a) and (b) show the Ricker wavelet and the coded signal used for the acoustic imaging, while Figs. (c) and (d) show the corresponding frequency responses.

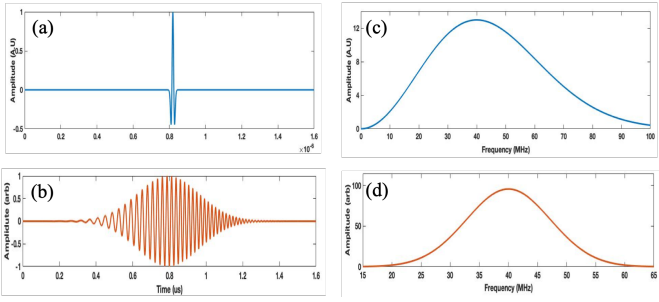

Fig. 1. (a) The Ricker wavelet and (b) chirped code used as transducer excitation pulses. The corresponding amplitude spectra are shown in (c) and (d).

## V. RESULTS AND DISCUSSION

A prepared slice of articular cartilage as shown in Fig. 1 was used as a test sample to explore differences between the investigated transducers. For each transducer both Ricker and chirp coded pulses were used as excitation pulses, and the recorded time series from these pulses were averaged 16 times to obtain some noise reduction.

The averaged backscattered signals were digitized into time series containing 2048 samples for each pixel in a *xy*-scan matrix, and saved to a HDF5 file format for further processing in MATLAB. This processing includes for example, wave compression of the coded waveforms which was done using a Wiener filter in the Fourier domain as previously described in [11]. After wave compression, the time series were finally processed into C-scan and B-scan images.

Examples of C-scans taken at a shallow depth into the cartilage sample is shown in Fig. 4. Here Figs. 4 (a) and (b) show the scans for the commercial and in-house transducers, respectively, driven by the Ricker pulse.

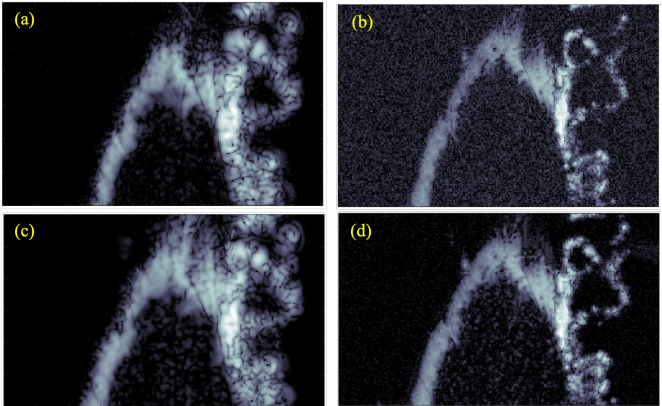

Fig. 4. C-scan images generated of the test sample using different transducers and excitation pulses. These are (a) commercial transducer with Ricker wavelet, (b) in-house transducer with Ricker wavelet, (c) commercial transducer with chirp code, and (d) in-house transducer with chirp code. All images have a dimension of 3× 5 mm$^2$.

The corresponding results when using a chirp excitation pulse are shown in Figs. 4 (c) and (d). B-scans of the test sample are shown in Fig. 5. with the (a) to (d) figures corresponding to the transducer-waveform combinations previous used Fig. 4. The red lines in Fig. 5 show the spatial locations were the C-scans in Fig. 4 were taken.

From Figs. 4 and 5 we notice e.g. by comparing the (a) and (b) figures, that the commercial transducer yields a significant higher SNR than the in-house transducer. Single pixel (salt & pepper) noise is for example, clearly visible in all domains of Fig. 4 (b) while it cannot be spotted in Fig. 4 (a). The observed difference in SNR is probably strongly related to the difference in transducer aperture sizes listed in Tab. 1.

The commercial transducer also explore more details from the inner parts of the cartilage, especially from the deepest parts. This is clearly seen e.g. by comparing the results for the chirp waves shown Fig. 5 (c) and Fig. 5 (d). Here Fig. 5 (c) shows elliptical shaped scatters originating from cells through the entire sample, while these scatters are visible only towards the upper surface in Fig. 5 (d). On the other hand, the elliptical scatters from the in-house transducer appear significant smaller than from the commercial transducer, suggesting that the point-spread function from the latter is larger. It is also possible to observe finer details on the C-scan from the in-house transducer, which together with a smaller point-spread function suggest that this transducer has a significant better resolution. This result can to some extend be explained by the smaller f-number of the in-house transducer.

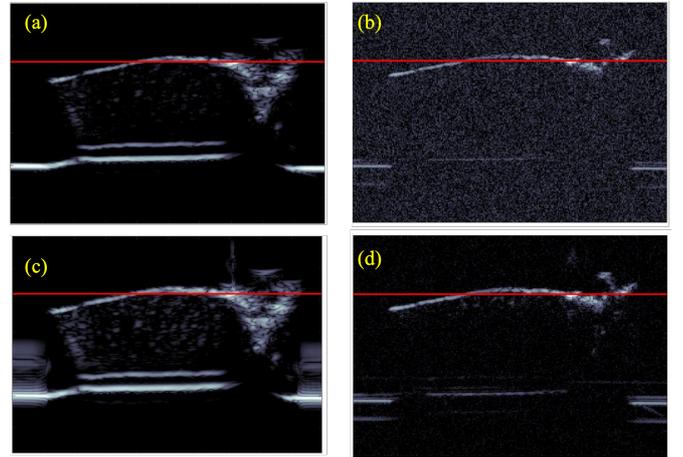

Fig. 5. B-scan images generated of the test sample using different transducers and excitation pulses. These are (a) commercial transducer with Ricker wavelet, (b) in-house transducer with Ricker wavelet, (c) commercial transducer with chirp code, and (d) in-house transducer with chirp code.

## VI. CONCLUSION

The results have demonstrated that the homemade P(VDF-TrFE) transducer is capable of producing images with good resolution of the cartilage test sample, and also visualising cell structures in the upper part of the sample. The in-house transducer also yielded significant better resolution and a smaller point-spread function than the commercial PVDF transducer. However, the SNR of the in-house transducer is currently not large enough to penetrate deep into the cartilage sample, and therefore not suitable for volume imaging of thick samples. The B- and C-scans taken with both transducers clearly show that the chirp-coded waveform is superior to the Ricker wavelet in terms of SNR and dynamical range. It is also believed that the SNR for the in-house transducer can be improved significantly by redesigning the systems pre-amplifier for the higher electrical impedance and the lower signal level produced by this transducer.

## ACKNOWLEDGEMENTS

We would like to thank Dr. Ashraful Islam for providing us the cartilage sample. A. Ranjan would like to give acknowledge Norwegian Ph. D Network on Nanotechnology for Microsystems for providing the travel grant for attending the IEEE conference.